# Normative Challenges of
# Risk Regulation of Artificial Intelligence
# and Automated Decision-Making

Carsten Orwat*[1], Jascha Bareis[1], Anja Folberth[2], Jutta Jahnel[1], Christian Wadephul[1]

[1] Karlsruhe Institute of Technology,
Institute for Technology Assessment and Systems Analysis

[2] University of Heidelberg, Institute of Political Science

*Corresponding author, e-mail: orwat@kit.edu

Preprint (under review)

November 2022

Karlsruhe

*Funding and acknowledgement:* The research within the project "Governance of and by algorithm" (GOAL) leading to the current paper received funding from the German Federal Ministry for Education and Research (Grant No. 01IS19020B). The funding is gratefully acknowledged. We would like to thank Karen Yeung, Ingrid Schneider, Catharina Rudschies, Armin Grunwald, Isabel Kusche, Reinhard Heil, Torsten Fleischer, Paul Grünke, Christina Timko, and Marc Hauer for their helpful comments.

# Normative Challenges of Risk Regulation of Artificial Intelligence and Automated Decision-Making

*Abstract:* Recent proposals aiming at regulating artificial intelligence (AI) and automated decision-making (ADM) suggest a particular form of risk regulation, i.e. a risk-based approach. The most salient example is the Artificial Intelligence Act (AIA) proposed by the European Commission. The article addresses challenges for adequate risk regulation that arise primarily from the specific type of risks involved, i.e. risks to the protection of fundamental rights and fundamental societal values. They result mainly from the normative ambiguity of the fundamental rights and societal values in interpreting, specifying or operationalising them for risk assessments. This is exemplified for (1) human dignity, (2) informational self-determination, data protection and privacy, (3) justice and fairness, and (4) the common good. Normative ambiguities require normative choices, which are distributed among different actors in the proposed AIA. Particularly critical normative choices are those of selecting normative conceptions for specifying risks, aggregating and quantifying risks including the use of metrics, balancing of value conflicts, setting levels of acceptable risks, and standardisation. To avoid a lack of democratic legitimacy and legal uncertainty, scientific and political debates are suggested.

# 1 Introduction

Recent proposals and approaches explicitly aimed at regulating artificial intelligence (AI) and automated decision-making (ADM) systems advocate risk governance or risk regulation using risk-based or risk-adequate approaches [1; 2; 3; 4; 5; 6; 7]. In general, the term risk regulation as umbrella term encompasses various approaches of "governmental interference with market or social processes to control potential adverse consequences" [8:3]. The notion of risk governance refers to complex arrangements of governmental, semi-private or private actors in identifying, assessing and managing or regulating risks [9; 10; 11]. Both risk regulation and risk governance aim to identify and inform about risks to society, decide how to balance risks and benefits, what the level of protection is or what level of risks is held acceptable for those affected and what measures should be taken to manage those risks, in particular, to minimise them.

The 'risk-based approach' is one of many types of risk regulation.[1] One of its meaning is to adjust or prioritise regulatory activities according to the risk levels attributed to the regulatory objects. This aims, in particular, not to overly stifle innovations and to save resources of the regulatory authority, e.g., by focusing efforts only on regulatory objects that are assigned a 'high risk' [10:330; 14; 15; 13; 16; 17]. This form of risk-based approach is most often proposed for regulating AI and ADM applications, including the proposed Artificial Intelligence Act[2] (hereinafter: proposed AIA). In contrast, another form of risk regulation, the rights-based approach, attempt to prevent risks by establishing regulatory rules that apply independent of the level of risks of the regulatory objects.[3]

Several studies[4] and official statements[5] emphasise the risks to fundamental rights and societal values caused by applications of AI and ADM, in particular on human rights and fundamental rights. The right to protection of human dignity, the right to life, the right to equal treatment and non-discrimination, the right to liberty, the right to free development of personality and autonomy, the right to informational self-determination, data protection and privacy, the right to freedom of thought, conscience and religion, the right to freedom of expression and information, the right to freedom of assembly and association, the right to a fair trial and to an effective remedy, the economic and social rights of fundamental rights provisions, the right to environmental protection and sustainable development, and the fundamental value of the common good are among the many fundamental rights and values addressed there.

---

[1] For the difference and relation between the 'risk-based approach' (or risk-based regulation) as a specific approach and risk regulation as the more general approach see [12:187; 13:508-510].
[2] European Commission (2021): Proposal 2021/0106 (COD) of the 21.4.2021 for a Regulation of the European Parliament and of the Council Laying Down Harmonised Rules on Artificial Intelligence (Artificial Intelligence Act) and Amending Certain Union Legislative Acts; Brussels: European Commission.
[3] For a discussion in the field of data protection see [17].
[4] In particular the studies of the Council of Europe [18; 19; 20; 21; 22; 3], the studies of the High Level Expert Group by the European Commission [23; 24], the study of the European Union Agency for Fundamental Rights [25], in Germany, the Data Ethics Commission [6] and the Enquete Kommission Künstliche Intelligenz [7]. See also [26; 27; 28; 29; 30].
[5] [5; 31; 2; 32].

The purpose of this article is to point to specific challenges of risk regulation with regard to AI and ADM applications that arise from the specific types of risks, i.e. risks to fundamental rights and societal values. This implies that many normative choices and value judgements are involved, which are the main subject of the following sections. In this paper, the term 'fundamental rights'[6] includes constitutional rights and the fundamental rights of the European Union. 'Fundamental societal values' are those values that are of public interest and considered essential for the functioning of a democratic society, but are not or not fully enshrined as rights in constitutions or the body of human rights. Examples for fundamental societal values are democratic structures and procedures, the common good or the striving for efficiency or innovation. The following argumentation is also based on the premise that the protection of fundamental rights is primarily the duty of the state [e.g., 33:38-41; 34:103; 35]. The duty to safeguard and guarantee them is thus also one of the main justifications for risk regulatory activities by state actors.

While there are numerous proposals for regulating AI and ADM, there is a lack of studies considering the specific characteristics of the risks to be regulated.[7] This paper aims to contribute to filling this gap, since the authors assume that an understanding of the specific risk characteristics is essential for establishing and maintaining a legitimate, effective and efficient risk regulation. To approach the manifold risk characteristics, the article has an interdisciplinary perspective, including political science, philosophy, law, risk research and technology assessment.

After shedding some light on the characteristics of AI and ADM applications as objects of risk regulation (Section 2), the European approach of an AI regulation (proposed AIA) is outlined in Section 3. In Section 4, the need for operationalisation of risks to fundamental rights and societal values is discussed. Normative ambiguities in risk regulation of AI and ADM are presented with four examples of fundamental rights and values in Section 5. From these examples we generalize and extend the discussion on normative ambiguities in Section 6 pointing to normative choices necessary to be made for a legitimate risk regulation. This leads to the question who should make such normative choices, which is addressed in Section 7, looking at the dispersion of normative decisions in the proposed AIA.

---

[6] Human rights are also referred to in specified places.
[7] Exceptions are [36; 37].

## 2  Applications of artificial intelligence and automated decision systems as objects of risk regulation

Although a commonly accepted definition of AI is still missing, we refer to an understanding of AI at the European policy level, in which AI systems include perception through data acquisition by sensors on the one hand and actuation with physical actuators or other forms of digital output (e.g., as provided information on websites and online platforms) on the other. Decision-making is based on knowledge representation or machine learning, trained on data sets and with automated inferences, and can adapt by analysing how the environment is affected by previous decisions made by the system [38; 39]. This narrows the scope of consideration of AI systems, as in most of the above proposals for risk regulations, to AI types and applications that are part of decisions about humans, i.e. as ADM applications. This decision-making is done either fully automatically by the system or semi-automatically by informing human decision-makers with recommendations by the system.

One of the areas of AI and ADM applications that raises considerable societal concern is their use for automated identification and differentiation of persons. Differentiation is mainly achieved by creating categories to sort persons as individuals or groups, by selecting individuals or groups or by assigning scores, ratings or rankings to them. This is done with criteria that are identified as relevant based on the discovery of correlations or the recognition of patterns in data sets with the help of machine learning system. It enables cost-effective, fine-grained targeting of groups or individuals with information about jobs, political statements, products, services etc. These practices are ushered in with the notions of 'micro-targeting', 'personalisation', 'individualisation', 'customization' or 'psychological targeting'. Other features of some AI applications are real-time inference and dynamic adaptation of decision rules through continuous analysis of data streams [40]. AI and ADM applications can have significantly expanded scale of impacts of decision-making ('mass phenomenon') because, unlike the limited range of decision-making by a single human decision-maker, ADM systems target all persons affected with the same set of decision rules and potential adverse effects [41:22]. This also implies that small biases or errors in data sets or algorithms can cause impacts on large portions of the population. Furthermore, in some AI and ADM application cases, feedback loops emerge by the use of algorithms to make decisions and by subsequently using information about these decisions to further train these algorithms.

The 'opacity' and 'black box' feature or incomprehensibility and unpredictability of AI and ADM applications is seen as another critical characteristic. It provides rationales for regulation, such as in Recital 47 of the proposed AIA (Explanatory Memorandum of the proposed AIA, p. 30, also in [42; 2]). However, not all algorithms and ADM systems are equally 'opaque' and unpredictable. It is important to differentiate whether incomprehensibility exists for developers and deployers or for external actors such as affected persons or oversight bodies. It is also argued that algorithms, even those of machine learning, can facilitate proving discrimination because programmed decision rules can be used as evidence [43]. However, many AI and ADM systems have a certain degree of incomprehensibility due to the 'learning' processes and the large number of variables, the complexity of relations between them (e.g., artificial neural networks) or the constant adaptation to data streams [44; 45]. Software testing and empirical investigations of outcomes are possible for developers and providers to detect biases and discriminations. Thus, the issue of the incomprehensibility of certain types of AI is not only an issue of technical complexity, but also of the amount of effort and resources allocated to increase comprehensibility. Second, the lack of external comprehensibility can also be due to the legal protection of proprietary software and trade and business secrets [46; 45]. Third, also the types of applications in business and administration with algorithm-based personalisation or individualisation of information, products, services, etc. lead to reduced comparability of individual treatments or outcomes with those of other persons.

These factors lead to situations with decreased comprehensibility of the decisions and outcomes, their underlying criteria and reasons leading to certain decisions and of the weighting between those criteria. This results in a diminished ability of those affected to detect, prove and contest adverse outcomes such as manipulations or discriminations [47; 48]. Therefore, the regulatory approach in the existent legal frameworks of anti-discrimination and (largely) data protection seems inadequate for the risks of AI and ADM. The latter regulatory approach is to provide legal support for affected persons, who has to become aware of adverse effects or violations of their rights or freedoms after they have occurred and to seek for redress and compensation. The difficulties to recognise and mitigate adverse impacts *ex post* are one of the main justifications for *ex ante* prevention with risk regulation.

## 3    Proposal of the European Artificial Intelligence Act

In a rough overview, the AIA proposal contains various instruments of risks regulation graded according to the four risk categories formed by the European Commission.[8] (1) AI practices with 'unacceptable risks' are prohibited, including certain manipulative techniques, certain social scoring practices, and under certain conditions some remote biometric systems (Article 5 proposed AIA). (2) For 'high-risk' AI systems the provider of an AI application has to fulfil several requirements. AI systems are categorised as 'high-risk' systems by listings in Annexes II and III of the regulation. In particular, the provider of a high-risk system has to establish and maintain a quality management system, including a documented risk management system, obey to data governance and management practices, create technical documentation and logging, manage human oversight, provide information to users (i.e. those who buy, deploy and use the system), build the system according to an appropriate level of accuracy, robustness and cybersecurity, perform a conformity assessment and, with passing this assessment, can use the CE mark for their systems (Title III proposed AIA). Additionally, the provider has to establish and maintain a post-market monitoring system spanning the lifetime of the AI application (Article 61 proposed AIA). (3) For AI systems rated 'limited risks' by the European Commission, the provider or the user have to obey to transparency obligations. AI systems of this risk category contain certain bots, certain systems for emotion recognition and biometric categorisation, and 'deep fake' systems. The transparency obligations are narrow and contain only information about the existence of the system, their operation or interaction or about that the content is manipulated (Article 52 proposed AIA). (4) For AI systems with 'minimal risk' there are no binding provisions, only the option of voluntary application of codes of conduct, which can be created by providers themselves or by organisations representing them (Article 69 proposed AIA).

---

[8] The AIA proposal has risks to health, safety and fundamental rights in its focus (e.g., Recital 1, 4, 5, or 13 proposed AIA), from which in the following only those to fundamental rights are considered.

The AIA has an approach of hybrid governance, in particular for high-risk AI systems. The aforementioned conformity assessments by most of providers of high-risks systems are a kind of self-assessment and self-certification. Also, the post-market monitoring by providers is an element of 'regulated self-regulation'. Beside this, the AIA proposal provides governmental supervisory and sanctioning activities by national supervisory authorities or market surveillance authorities (MSAs). For MSAs, the AIA proposal provides several regulatory means, including to gain access to data and the source code, evaluate AI systems, require corrective measures, prohibit AI systems, withdraw or recall them (Article 63-68 proposed AIA).

## 4  Necessity for operationalising risks and its challenges

The effectiveness of risk regulation is assessed in view of the actual achievement of protection goals, such as prevention of damage, for which regulators must provide accountability [10:332-336]. Sufficiently unambiguous and concrete criteria or principles for what constitutes relevant risks are necessary not only to avoid arbitrary assessments, but also to provide legal certainty in risk assessments or compliance tests and to derive the appropriate level of regulatory measures such as tests, approvals, requirements, bans or moratoria. Regulators themselves can use such criteria and principles to avoid objections and legal actions due to allegedly incorrect or inaccurate assessments. Last but not least, clear risk definitions also provide developers and operators with the necessary orientation in the anticipatory and preventive development, design and application of systems and, thus, security in financial investments. However, potential problems with inappropriate selection and prioritisation of protection goals and risks as well as their inappropriate interpretation and operationalisation can, in the worst case, result in the actual protection goals of regulation not being achieved.

This raises questions how to get to the necessary interpretations, specifications and operationalisations. It is questionable whether the now numerous ethical guidelines in the context of AI and ADM[9] can be used directly for risk operationalisation in risk regulation. Since their non-binding character, they can only provide inspiration for identifying and interpreting values affected by AI and ADM applications. Some of them do not address European or national specificities, in particular, the historical development of certain constitutional rights and their position in the value structure of a society. Moreover, they differ greatly in terms of the values and principles included. The respective selection of values and principles and their specific interpretations are often non-transparent and difficult to reconstruct, as are the underlying normative perspectives or interests. The question arises of whether the requirements of some guidelines actually fall short of existing legal requirements and thus are no more than an 'ethics-washing' [52].

Also, the existing secondary legislation framework provides only few points of reference for interpretation and operationalisation. For instance, the General Data Protection Regulation[10] (GDPR) and the German General Equal Treatment Act[11], which are relevant to AI and ADM particularly in the national context due to their regulatory scope, are not only considered inadequate to protect against the risks of AI and ADM [18; 3:21-25, 35; 53:128-142]. Furthermore, they are often itself compromise texts that serve to balance interests of different stakeholders. They do not provide 'pure' interpretations of fundamental rights. Additionally, they often give too much room in interpretations for an *ex ante* specification of risks, because they are often designed for *ex post*, context-dependent judicial decisions that require interpretations by judges [in general, 54:13]. Such legal rules for contextual judicial decisions are difficult to transform into generalised standards for the purposes of an *ex ante* oriented risk regulation.

---

[9] Overviews are provided by [49; 50; 51].

[10] Regulation (EU) 2016/679 of the European Parliament and of the Council of 27 April 2016 on the protection of natural persons with regard to the processing of personal data and on the free movement of such data, and repealing Directive 95/46/EC (General Data Protection Regulation) [2016] OJ L119/1.

[11] General Act on Equal Treatment of 14 August 2006 (Federal Law Gazette I p. 1897), as last amended by Article 8 of the SEPA Accompanying Act of 3 April 2013 (Federal Law Gazette I p. 610), in German: Allgemeines Gleichbehandlungsgesetz vom 14. August 2006 (BGBl. I S. 1897), das zuletzt durch Artikel 8 des SEPA-Begleitgesetz vom 3. April 2013 (BGBl. I S. 610) geändert worden ist.

Rather, the practices and case-law of the protection of fundamental rights, as well as their broad and well-founded scientific debate and further development, can inform the interpretation of their protection goals, essence and options for specification and operationalisation [55; 56; 23; 57:82-83]. However, despite their extensive elaboration in the international body of fundamental rights, also fundamental rights are not always unambiguous in their content and are not consistently established and accepted in different jurisdictions [e.g., 33:35-52; 58:20; 36]. In addition, interpretations and operationalisations of fundamental rights need to be continuously adapted to the rapid socio-technical developments in AI and ADM (see Section 6.4). Likewise, the comprehensive ethical research on AI and ADM provides valuable guidance for identifying and operationalising risks. However, the challenge with the latter is to deal with the plurality of their approaches and worldviews. These approaches, legal and ethical, have yet to be linked to the interpretational needs and operationalisations for the normative basis of risk regulation of AI and ADM applications. Normative decisions are required to determine which of the many possible operationalisations are acceptable.

## 5   Examples of normative ambiguities

The notion of normative ambiguity refers to differences in the understanding and acceptance by actors of risk regulation of the meaningful and legitimate values, concepts, priorities, assumptions, or boundaries to be applied in risk appraisals [9:77f., 153; 59; 60]. As a result, there can be no clear evaluation criteria. Value concepts for the evaluation of risks can even conflict with each other. In the following, four examples of fundamental rights and societal values affected by AI and ADM, considered from a European and German perspective, are used to show that the operationalisation or concretisation required for risk regulation is challenged by normative ambiguities. Human dignity is considered the most central fundamental right. The rights to informational self-determination, data protection and privacy as well as non-discrimination, equality, justice and fairness are the most discussed rights affected by AI and ADM. The societal value of the common good is considered here because of its peculiarity of going beyond the protection of individual rights and freedoms.

### 5.1   Human dignity

AI and ADM applications can affect the inviolable right to human dignity [61; 23:10; 6:43; 19:33-34; 3:27-28; 25:60; 1: Article 5; 62] (see also Recitals 15, 17, 28 and 37 proposed AIA). Individuals are evaluated, judged and differentiated based on a single, a few or a multitude of numerical values. Most machine learning applications use data about statistically generated groups of people and about the past to build models for algorithmic decision-making using classifications, scores or predictions. In many cases, they use correlations and constructions of persona types, such as in machine learning methods for predicting credit worthiness, likelihood of recidivism, potential suitability as an employee etc. These data analyses do not take account of the individuality of persons and their unique individual qualities. Instead, they sort individuals into prefabricated categories of group characteristics based on single or multiple variables. A violation of human dignity can result from the treatment of persons as mere means, instruments or objects. The so-called 'object formula' prohibits reducing individuals and treating them as mere objects to achieve the goals of others, because individuals have an intrinsic moral worth.

However, controversies over the interpretation of this fundamental right point to necessary normative choices and clarifications that a society must make with regard to AI and ADM applications. At issue are the questions of what it means that the right to dignity is inviolable and absolute, as enshrined in the international and national body of human and fundamental rights law.[12] A question is what precise specifications follow from this norm to determine when it has been violated. Conceptual features for operationalising human dignity for assessing the risks of AI and ADM applications have not yet been worked out. Crucial aspects may be, for example, specifications of the object formula, the prohibition of instrumentalisation and reduction of humans, conditions for the affected individual's consent to a certain treatment, for the existence of contempt or humiliation, for manipulation, for ensuring the necessary respect for the individual's subject qualities [65], for humans' ability to act in a self-determined manner, for the free personal development, or the possibility of leading an autonomous life [e.g., 66].

Although the treatment of individuals as numbers is inherent to information and communication technologies with personal data processing, also when used in decision-making by humans, a distinctive feature of fully automated decision-making is that usually no further information about the person affected and their situation is considered. Assessments and judgments about individuals are based on numerical values derived from processing data about large numbers of persons categorised into specific groups. Any judgment of individuals based only on numerical values derived from processing of data about groupings is a normative choice between efficiency gains on the one hand and recognition of the individual and their uniqueness and dignity on the other. Furthermore, algorithms cannot give reasons and justifications for the decisions taken, which is important if those decisions shall be contested in cases they violate human dignity. Humans make decisions based on hermeneutical processes and not like decisions based on machine learning with the use of data [54]. Problematically, data does not speak or justify itself, it needs to be interpreted and normatively justified by humans. This is one of the reasons why algorithmic decisions on differentiating people are more problematic than human-based decisions.

---

[12] In particular, Article 1 of the Charter of Fundamental Rights of the European Union (hereinafter Charter) and Article 1 (1) German Basic Law. However, internationally, also in Europe, human dignity is institutionalized differently in constitutions, has different ways and degrees of specification and different positions in the value structures of the respective societies [63; 64].

Another phenomenon is the potential opacity of the rules, criteria and their respective weightings that determine decisions in ADM, as discussed above. This touches on a crucial dimension of dignity, namely the consent of persons to the treatment, since persons affected may not know what they agree to [67], especially the unknown criteria of decisions. Furthermore, Karen Yeung points out the problem that economic operators of algorithmic differentiations primarily aim to achieve economic value from customer relationships instead of elaborating the actual rationales of their behaviour [19:30f.]. Another potential factor pointing to violations of human dignity are comprehensive personality profiles or 'super scores' that imply 'total' surveillance of persons [6:43]. However, profiles and scores that are transferred and merged between organisations are already part of many business and government practices. Therefore, normative decisions on the tipping points between justified and unjustified comprehensive profiles are needed for specific contexts.

Secondary law does not provide sufficient guidance for operationalising human dignity. The proposed AIA prohibits certain types of social scoring, manipulative systems, and certain biometric system applications (Article 5 proposed AIA). These prohibitions are justified, among other things, with the protection of human dignity (Recital 15 and 17 proposed AIA). However, the provisions of the proposed AIA are criticized that the criteria that qualify the AI systems relevant for the prohibitions would limit the scope of the prohibition so much that they could turn out to be meaningless [68:11]. Additionally, the need to develop criteria for distortions of human dignity is particularly relevant for the 'high risk' AI applications and their residual risks.

In data protection law, in particular in Article 22 GDPR, the prohibition of exclusively automated decision-making originally has the intention to ensure protection from being treated as a mere object [69:GDPR Article 22 point 3]. Due to several exceptions in the GDPR, the Article cannot be seen as a general prohibition. On the contrary, it is rather a regulation of the conditions under which fully automated decision-making is allowed. Unclear provisions on the conditions under which a human is and must be involved in decision-making or on the obligation of the data controller to explain the 'logic involved' raise doubts whether, overall, human dignity is sufficiently protected by the provisions of the GDPR.

The specific operationalisation of the right to dignity should therefore be substantiated by *ex ante* definitions of the conditions under which the prohibition of ADM should take effect. This could be particularly relevant for cases where the different criteria used by ADM and the weighting relationships between them cannot be explained by the deployer or user to the affected persons, or where specific goods or positions that are essential for leading a decent life and for the free development of personality are the objects of ADM.

**5.2  Informational self-determination, data protection and privacy**

AI techniques can continue and accelerate trends of data mining, big data analytics, predictive analytics, profiling and targeted advertising that are considered problematic for data protection and privacy. AI techniques increase capabilities of data aggregation, data repurposing, de-anonymization and re-identification of persons even from non-personal or anonymized data, as well as capabilities to assess, categorise, rank, or score persons. Further capabilities of AI are to make automated inferences about the identity, personality characteristics, and other sensitive facts also from rather 'innocuous' or mundane forms of data (e.g., communication in social networks), or to provide algorithms and models in ADM that affect persons' abilities and preconditions to self-develop and lead a decent life. AI systems, furthermore, increasingly intrude into the most intimate sphere of individuals (e.g., AI-based 'personal assistants'). Since AI-based biometric identification technologies, including facial recognition systems or emotion recognition systems, exhibit multiple of these capabilities they are particularly problematic [e.g., 19; 26; 25].

Although the fundamental rights to data protection and privacy and to informational self-determination have been established for decades, there is still a certain degree of normative ambiguity. These fundamental rights can persistently be interpreted differently, among other things, with the reason to be adaptable to new socio-technological developments. At the European level, the rights to the protection of personal data and to privacy are particularly enshrined in Article 7 and 8 of the Charter, Article 16 (1) of the Treaty on the Functioning of the European Union (TFEU), Article 8 of the European Convention on Human Rights (ECHR) and the Council of Europe Convention for the Protection of Individuals with regard to Processing of Personal Data (Conventions 108 and 108+). The rights are further developed by case law by the Court of Justice of the EU (CJEU) and the European Court of Human Rights (ECtHR).[13] At the German national level, for instance, the right to informational self-determination has a status of a fundamental right. It is based on the German Basic Law, Article 2 (1) on the free development of personality in conjunction with Article 1 (1) on human dignity.[14] The right to informational self-determination has been often understood as control over personal data by the individuals or data subjects affected by the data gathering and processing. The fundamental right is conceptually implemented in data protection law, in particular, by the GDPR mostly through rights for the affected individuals to be informed, to be asked for consent, to rectify or erase data, to restrict data processing, or object fully automated decisions (Articles 12-22 GDPR). However, the GDPR itself causes contradictions with options to control by the persons affected, because, in particular, it allows, under certain circumstances, data controllers to process personal data according to their legitimate interest without the consent of data subjects (Article 6 (1) f GDPR).

---

[13]  De Terwangne [70] describes that the Conventions 108 and 108+ as well as the case law by the European Court of Human Rights (ECtHR) has developed data protection as right to informational self-determination with a right to control. For the rights to privacy and data protection, Brkan [71] shows for the case law of the Court of Justice of the European Union (CJEU) that until then the court has not provided a clear normative framework for the essence of both rights. Fuster and Gutwirth [72] points to contrasting interpretations of the fundamental right to data protection of the Charter with respect to the interpretation as informational self-determination either as a prohibitive notion or as permissive (or regulatory) notion.

Besides the interpretation as 'data control', another interpretation of the right to informational self-determination addresses more directly the protection of human dignity and the free development of personality (see Decision BVerfGE 65,1 pp. 42ff.) [73; 74]. This interpretation calls for the preservation of largely self-determined options for action in informational contexts as well as the free formation of identity, which should still largely be perceived as one's own. Furthermore, it demands ensuring the uninhibited use of digital services and products and avoiding chilling effects that can emerge from uncertainties. These uncertainties arise from data processing and its consequences, which are no longer traceable and understandable for the persons concerned (Decision BVerfGE 65,1 pp. 42ff.). Among other things, the latter should prevent impediments to political participation and freedom of expression. In particular, this interpretation can justify stronger regulation of algorithm- and data-based decisions or ADM. Risk assessments in regulatory approaches would then not only need to address the possible loss of control over personal data, but also the possible violations of human dignity, restrictions on options of free self-determination, of the right to self-representation, options to influence decisions by the persons affected, and to form one's own identity. These aspects would need to be substantiated based on specific criteria and principles.

In general, there are many different (theoretical) concepts of (informational) privacy [e.g., 75; 76; 77; 13:537-539], from which the rights to informational self-determination, data protection and privacy have also been developed. A consequence of this diversity of conceptions is that there is no clearly delineated and generally accepted list of adverse impacts or risks to the rights of informational self-determination, data protection and privacy.[15] The adverse impacts comprise various forms of restrictions of freedoms, including restriction of the above-mentioned freedoms of action and self-determination in development of personality, identity formation, as well as adverse impacts through chilling effects. In addition, since data protection also serves to prevent discrimination, the potential violating impacts include unjustified unequal treatments as well as unjust attributions of individual characteristics to persons. Associated with this are adverse impacts such as violation of human dignity, stigmatisation, stereotyping, damage to reputation, abuse of information power and structural superiority. Other adverse impacts include conformity pressures through surveillance, identity theft, disappointment of confidentiality expectations or persistent risks due to the permanence of data storage and data processing [e.g., 78; 79; 13:537].

---

[14] For AI, the protection of the right to informational self-determination is demanded by Bundesregierung [32:10, 16, 29 etc.] and Council of Europe - European Committee of Ministers [5:Appendix Point B.2.1]. The right is established and developed by a series of decisions by the German Federal Constitutional Court, such as BVerfGE 65, 1 ('Volkszählung', judgement of 15 Dec 1983), BVerfGE 113, 29 ('Anwaltsdaten', judgement of 12 Apr 2005), BVerfGE 115, 320 ('Rasterverhandung II', judgement of 4 Apr 2006), or BVerfGE 118, 168 ('Kontostammdaten', judgement of 16 Jun 2007).

Another consequence of the conceptual diversity is that attempts to quantify these fundamental rights have different underlying worldviews about what data protection and privacy are for and what and how to protect them. In particular, recent approaches to quantify and measure privacy with 'privacy metrics' [e.g., 80], such as 'k-anonymity' or 'differential privacy', are based on a narrow understanding of privacy as anonymity, secrecy or confidentiality of systems. This is an inappropriate reduction, because the abovementioned protective goals of the rights to informational self-protection and data protection go far beyond this narrow interpretation of privacy [81].

## 5.3 Justice, fairness and anti-discrimination

A third example of normative ambiguity relates to the concepts of anti-discrimination, fairness and justice.[16] According to the existent anti-discrimination law, AI and ADM applications cause a variety of discrimination risks. The main causes of these risks include cognitive biases, technical or organizational errors, or subjectivity in decisions of developing, adapting or using the systems, including the use of biased or non-representative data sets for machine learning, inappropriate data labelling, or inappropriate selection of AI algorithms or parameters. In sum, they can lead to inappropriate use of attributes protected in anti-discrimination law (e.g., gender, age, or religion, also called grounds of prohibited discrimination) or of proxies related to these protected attributes [82; 83; 18; 47; 41; 84].

---

[15] The wide range of adverse impacts is also illustrated by enumerations in Recitals 75 and 85 GDPR, but are not completely and exhaustively defined there. The openness of the range of impacts, enables to potentially include new types of adverse impacts by socio-technological developments, but leaves a certain degree of uncertainty in *ex ante* determining risks factors that have to be considered in risk assessment and risk management.

[16] At the European fundamental rights level, the rights to equal treatment and non-discrimination are enshrined in Article 20, 21 and 23 of the Charter, Article 2 of the Treaty on European Union (TEU), Article 8 and 10 of the TFEU and Article 14 of the ECHR and Protocol 12.

For risk assessments, anti-discrimination law like the German General Equal Treatment Act can only provide a rough basis and scope for concretisation, since the Act itself leaves room for interpretation and ambiguity [41:74-76]. It is a general question, whether those context-independent rules for less discriminating algorithm-based differentiations can be defined at a general level at all [see also 48]. The principle of necessity and proportionality, that is underlying anti-discrimination law, implies rather context-specific balancing decisions about the legitimacy of a differentiation considering factors such as purposes and contexts of types of differentiation, the legitimate aims, necessity and appropriateness of the differentiation or the availability of less adverse alternatives.

Contrary to this open question, researchers and developers strive to define and quantify the discriminatory risks of machine learning algorithms with so-called 'fairness definitions', 'fairness metrics', or 'fairness measures' [overviews in 85; 84]. However, conceptions of fairness definitions are predicated on highly different normative worldviews, assumptions and ideas of justice and fairness [86; 87; 88; 89; 48]. Some fairness metrics merely depict relations between error rates, including false positives and false negatives.[17] Here, the danger of inflicting injustice is still looming, so those measurements of 'fairness' must be understood as expressions of residual risks (i.e. those risks remaining after preventive measures).

Furthermore, the use of such fairness metrics already takes for granted the application of ADM and decision criteria derived from generalisations. However, this assumption must be contested from the outset, since the legitimacy of the applications themselves can be questionable. Anti-discrimination law is not only based on the constitutional rights to justice and equal treatment, but also serves human dignity and the free development of personality by avoiding stigmatisation and unjustified attribution of characteristics to persons and their negative consequences for those affected. Many fairness definitions aim at the fair treatment of groups and do not consider individual justice, individualized justice or justice through case-by-case judgements ('Einzelfallgerechtigkeit'). Individual justice takes account of individual persons and their personal qualities and life situation (e.g., by personal interviews), which is usually abandoned when using fully automated decision-making. ADM and AI-based decision rules are usually based on generalisations from data about groups or entire populations and the use of detected patterns as decision rules in ADM applications [79; 90:519].[18] The problem is also demonstrated by one of the first judicial discrimination cases in Europe, in which the Court prohibited not only the exclusive use of protected attributes in automated credit decisions. It also emphasised the inadequacy of using statistical figures based on data collected on other individuals to assess the creditworthiness of the considered individual (Decision of the National Non-Discrimination and Equality Tribunal of Finland, [91]). Linking this perspective to the larger debate on justice, already allowing the use of AI and automated decision systems for specific contexts itself is a normative decision. Normative choices also encompass determining for specific contexts what rules, criteria or parameters may or may not be legitimately used in automated decisions and for which purposes, how persons affected can communicate their views or which options they have to challenge and redress automated decisions.

---

[17] For instance, the fairness metrics 'equalized odds', 'demographic parity' (or statistical parity), 'equal opportunity', or 'treatment equality' are different ratios built of false positives, false negatives, true positives, or true negatives rates [85; 84].

If one wants to operationalise risks of AI and ADM applications for social justice, equity and equality, a number of normative issues would have to be decided beforehand by society itself, because the use of algorithm-based differentiation and ADM inevitably has an impact on justice, equity and equality. However, there are several concepts of justice and equality.[19] This reflects not only the plurality and controversies in contemporary societies in this respect but also that different concepts are prevalent in different areas of societies. Therefore, it is necessary to identify the contexts in which these fundamental rights are affected and how, and which respective concept of justice should be applied in this particular context. In particular, concepts of justice and equality differ in their objects of consideration, e.g. respect for equal moral worth and dignity and valuing of individuals as equals, equality of rights and duties, equality of the distribution of welfare, resources and opportunities, human capabilities, or political and democratic status. They also differ in the conditions that exist in a society to recognise and overcome inequalities in order to achieve social solidarity and respect among its members. In particular, the questions under what conditions and for whom possible deviations from equality are justified is one of the most controversial political issues. Furthermore, specific conceptions use different operationalisation approaches, such as emphasising procedural justice (e.g., due process, right to remedy, presumption of innocence) or substantive justice with a focus on the outcomes of decision-making. The conceptions also differ in their implications for state interventions derived from equality rationales, such as compensating for disadvantage (and maintaining social structures in place), ensuring opportunities for a decent life or eradicating social structures seen as unjust or oppressive [53:219-226; 86; 87; 92].

---

[18] Here, the discussion about the risk of injustice from AI-based ADM applications extends and intensifies the discussion on statistical discrimination [86].

[19] The complexity of the justice discourse is reflected in the wide variety of justice theories and concepts, ranging from liberal and communitarian traditions to different (cap)ability approaches or more specific discussions about equality of opportunity, gender equality or intergenerational equity, to list but a few of the social science approaches.

## 5.4 The common good

A fourth and final example of a normative perspective that lacks clarity is the common good. Recent political strategy papers emphasise that AI applications should not only be viewed from the perspective of the individual, but should also contribute to the common good [23:4f.; 32:7, 9, 10, 45, 47; 2:2]. However, the concept of the common good is one of the most contested and vague. Following a liberal tradition, the common good conception embodies a shared standpoint for practical reasoning among the members of society that urges us to engage in "a way of thinking and acting that constitutes the appropriate form of mutual concern" [93:Section 4.1]. This concern raises awareness that there should exist some inclusive facilities, institutions, collective or public goods that the community has an obligation to maintain, such as public education, local transport, health care or energy infrastructure. The strength of the concept of the common good lies here in the recognition that individual rights, such as the free development of personality, also stand in a dependent relationship to this communal realm [see also 94:487]. Collective goods have an enabling function for individuals to become self-determined and autonomous selves.

In the context of ADM and AI, it is still unclear how this broad concept of the common good could be translated into concrete principles, criteria for risk assessments or concepts of systemic risks. There are a number of possible concretisations: (1) These include adverse outcomes for entire groupings of persons or at the collective level that are not covered by the statutory data protection and anti-discrimination framework, which is mainly focused on the individual [e.g., 95]. Adverse outcomes at the macro level may also result from a loss of trust or confidence, intimidation and chilling effects. These can be caused by abstract uncertainty about the processing of personal data or its transfer to other parties, about belonging to artificially fabricated groupings, and uncertainty about the actual criteria operating in algorithm-based decisions and determining their consequences. These uncertainties can cause certain population groups to withdraw from using digital services and products. This can lead to social segregation or damage processes of democratic decision-making, as digital services and products increasingly serve community building and political exchange. (2) Furthermore, as AI and ADM contribute to the fine-grained differentiation, personalisation and individualisation of services and products, they may displace practices based on solidarity or community services, changing the provision of a broad range of public or collective goods (e.g., health services). (3) New models for using data and AI methods, especially as open data concepts in public administration to support social innovations [7:198f., 201, 206f., 215] or as AI-enhanced platforms to optimise public and multimodal transport (ibid. 385f.), are also discussed in view of serving the common good. (4) The mitigation of substitutional effects, job losses or surveillance issues caused by AI-based systems in the area of work are debated as well with regard to the common good (ibid. 140f.). (5) Last but not least, the normative question of the societal distribution of risks, the distributive effects of feedback loops, as well as the distribution of efficiency gains and other benefits from AI and ADM can be approached from a perspective of the common good. All the above issues require normative choices for establishing risk regulation if the common good is to be addressed as a relevant value for risk assessments.

## 5.5 Interim conclusions

The above described examples demonstrate sources of normative ambiguities. First, they may result from a certain degree of interpretative openness and indeterminacy of abstract fundamental rights. The openness for interpretation and specification is required for fundamental rights to be applicable to different contexts and situations and in different times, including to new socio-technical developments. Second, the existent secondary law is often considered inadequate to protect fundamental rights from the risks of AI and ADM. Therefore, it cannot likewise be taken as orientation for specifying fundamental rights. Third, some fundamental rights and societal values can be endangered by multiple potentially affecting activities. Normative ambiguities can result from uncertainties and subjectivity in selecting of and weighting between relevant risk factors. Fourth, for some fundamental societal values, such as the common good, there is not enough research, consolidated knowledge and, in particular, public discourses and consensus to select appropriate interpretations and specifications for the risk regulation of AI and ADM applications.

# 6 Normative choices

Risk regulation, including risk assessment and risk management, usually involves multiple normative choices in particular to solve normative ambiguities [96; 97; 98; 99; 100; 101; 102; 103]. They range from the definition and interpretation of regulatory protection goals, to the determination of what constitutes a risk, and to acceptable levels of risk imposed on society. In the following, the numerous points of normative choices in risk regulation of AI and ADM are identified and discussed, including how with the proposed AIA the regulator intends to resolve normative ambiguities or lets them open.

## 6.1 Choice among risk-regulatory approaches

Already the application of the risk-based approach itself is a normative decision. Prioritising some risks, excluding others, and selecting areas of intervention as objects of risk regulation entails the danger that this will violate state authorities' guarantee of human rights protection for everyone. This happens by neglecting risks not assessed as 'high risk' but that are still relevant. The dismissal of this alternative option is a deliberate decision to allow a certain degree of risks of violating fundamental rights to the benefit of other societal values like innovation or efficiency gains. This implies that the chosen approach will not protect everyone equally, in particular by neglecting cases of 'non-high' but still relevant risk and, thus, leads to an unequal treatment of human rights holders [similar 34:102]. Another option of risk regulation would be to establish rights, duties, principles or other rules for risk prevention independent of specific risk levels assessed for AI and ADM applications.[20] For the proposed AIA, this regulatory option is deliberately dismissed by the European Commission citing administrative costs and burdens on regulatees, even though the Commission has found a more effective enforcement of existing laws on fundamental rights through this option [104:64-85]. Science and technology studies have pointed out that the designing of risk regulation schemes is influenced by narratives, framings and cultural contexts [e.g., 99]. For the proposed AIA, the narrative of choosing the specific form of risk-based approach seems to be that some degree of risks in form of potential derogations of fundamental rights is worth of saving administrative resources, not hindering innovation, or profits by gains in efficiency.

The choice between a horizontal or sector-specific regulatory approach is also a normative decision about the regulatory approach. The AIA proposal follows a horizontal approach attempting to establish the same rules across different sectors.[21] A cross-sectoral scheme using single interpretations of fundamental rights and values (e.g., a unique interpretation of justice or fairness) can negate that there can be different conceptions in different areas of society. For instance, a single definition of fairness would mean that one conception of justice is relevant to all AI and ADM applications in different contexts. However, this would negate that different conceptions of justice and fairness can be relevant to specific contexts, taking account of their peculiarities. For example, over decades of societal developments, the concept of equality of opportunity has become more relevant in certain contexts (e.g., work, education), while that of equality of legal and political status has gained importance in other areas (e.g., democratic processes) [similar 86:6f.].

## 6.2 Selection, prioritisation and operationalisation of risks

When a risk-based approach is chosen, further normative decisions are the selection and prioritisation of certain fundamental rights and societal values or protection goals (for risk assessments in general, see [101; 105]). This includes decisions about which risks to include or omit in risk assessments of risk regulation, how to weight between different risk types and adversely affecting activities, as well as the regulative measures (bans, required organisational measures, etc.) linked to different risk levels or categories [in general 96]. There is a risk that certain types of risk factors or adverse activities are ignored, in particular, with a lack of detailed guidance and open list of risks to certain fundamental rights and societal values. For example, it is questionable if the value 'common good' (or parts of it) is addressed as a protection goal by the proposal AIA regulation. For instance, the proposal does not include the consideration of systemic risks like chilling effects or the societal distribution of risks and benefits.

---

[20] For a discussion of the rights-based and risk-based approach in data protection law see [17].

[21] An exemption are financial services for which special provisions are made in the AIA proposal.

The abovementioned examples of normative ambiguities demonstrate that the interpretation and operationalisation of fundamental rights and societal values are one of the central normative decisions that a society have to made. These are choices among the possible concepts that interprets fundamental rights and values. The decisions include the adaptation of the concepts to the contexts of AI and ADM applications and result in operationalisations of regulatory requirements. The examples of possible implementation of conceptions of justice and fairness or the common good demonstrate that these seemingly practical matters of risk regulation are actually fundamental decisions about determinations on how to live together in a society.

## 6.3 Aggregation, comparison and quantification of risks

In general, several concepts of risks exist, which contain attempts to aggregate, quantify and measure risks as well as qualitative risk assessments. They are subject to long-lasting debates in science and risk governance about the appropriateness of the use of quantitative or qualitative concepts for specific regulatory purposes [96; 97; 9:12-45; 98; 103].

The AIA proposal also provides for the use of metrics (in particular Article 9 (7) or Article 15 (2) proposed AIA). Such metrics can include fairness metrics, privacy metrics, explainability metrics [106], or performance metrics such as accuracy metrics [e.g., 107:101ff.]. Motivations for regulators, developers, providers or deployers to strive for quantification can be to aggregate multiple dimensions of risk assessments into a few better manageable numbers or a single number. Numerical values ideally enable to set threshold values or quantitative targets for acceptable risk levels or to justify the sorting of systems to certain risk categories according to certain threshold values, scores or rankings. Quantitative measures would also ease conformity assessments, certification procedures, inspections of the quality, and the continuous monitoring of the use of AI and ADM applications over their lifetime. Even automated checks of reaching certain prescribed thresholds are conceivable. Numbers would enhance the comparison of the riskiness of alternative system designs and their optimization in development processes.

However, the aggregation and quantification of risks to fundamental rights has several problems and limitations. First, from science and technology studies it is known that aggregation and quantification can disguise the underlying normativity of the decision situation and can lead to inappropriate simplifications. Aggregation and quantification can create an "aura of objective truth" [108:1] and scientific neutrality, but in fact include several subjective assumptions, value judgements and political decisions [96; 99; 109; 108; 110]. Furthermore, the regulatory regimes and the problems, which are intended to be governed, are co-produced [111:422; 112]. The developing of a measurement or standardised scale shapes the way the world is experienced and co-produces the phenomenon it claims to measure [109; 108:12, 28]. Establishing a certain metric for AI and ADM applications can co-produce a specific understanding of dignity, privacy, justice and other fundamental rights and societal values.

Second, single AI and ADM applications usually have risks to several fundamental rights and societal values at once. The example of data protection and privacy shows that the fundamental right can be violated by a multitude of affecting activities. Selecting certain rights, values and protection goals or parts of them, or certain impacts to be considered for aggregation, for a measurement instrument or for commensuration (i.e. making comparable on a common measurement) are normative decisions of weighting between them. This includes choosing certain importance weights or forming risk categories (e.g., 'high risk' or 'low risk', or 'traffic lights' approaches of qualitative risk assessment). As exemplified above, selecting one privacy indicator or metric would mean that this can lead to an inadmissible simplification of the multiple protection goals of this fundamental right. Furthermore, the example of fairness metrics demonstrates that individual metrics are each based on their own and highly different understandings of fairness or justice. Selecting one fairness metric to be applied in risk regulation is a value judgement for one justice concept with a far-reaching effect for the entire society, also meaning the devaluation of alternative value concepts und worldviews. Accepting one or a few metrics suggest an unambiguity in the normative foundations that might not be there.

Third, a further limitation results from the context-dependency of such value judgements [97]. It hinders the use of common quantitative measures applied to all areas of society. For example, the decision between reducing the false positives rate or false negatives rate is a normative decision that usually differs in different areas of society. While, for example, in medicine it is usually more important to avoid false negatives (wrongly not detected ill persons), in criminal law it is usually more important to avoid false positives (wrongly convicted innocent persons) [102:124f.].

Fourth, quantitative measures can raise false expectations about the protection of fundamental rights. Metrics often provide a scale or range of numerical values. This may surrogate that fundamental rights can be partially derogated (such as to a certain percentage). Instead, fundamental rights have the status of universal moral boundaries that can only be deviated from in narrowly defined circumstances (see Section 6.4). This makes it problematic to rank fundamental rights violations "on a sliding scale from trivial to serious." [113:10]. As mentioned above, many fairness metrics use error rates that have to be understood as residual risks that still can lead to violations of fundamental rights. In concrete terms, this implies, for instance, that a certain percentage of persons affected may be discriminated as outcome of the application of AI and ADM systems even when their conformity with AIA provisions is self-certified.

Fifth, many risks to fundamental rights are deliberately not or notoriously difficult to mathematically formalise, quantitatively estimate, to measure or to aggregate. This is due to the characteristics of the fundamental rights. Fundamental rights and societal values are often hardly comparable, nor are they commensurable. For instance, violations of human dignity or restrictions on freedoms are hardly assessible in quantitative terms. It belongs to the ethical nature of human dignity that humans should not be evaluated in quantitative terms, but should only be evaluated in their unique individual qualities. Furthermore, some human rights, such as the inviolable right to human dignity, should actually not be subject to derogations or limitations and should not be balanced against other fundamental rights.

Sixth, the above examples of normative values also indicate that fundamental rights are complex value structures in the sense that they can have supportive or instrumental interrelations among themselves. Operationalising specific risks requires understanding the complex value structures with their supportive and derivative relations and acknowledging that the meaning of each fundamental right depends on other fundamental rights. This limits the possibility of operationalising one risk alone, such as the right to privacy only as mere technical anonymity or the right to anti-discrimination solely as 'fair' distribution of outcomes.

Seventh, building quantitative measures causes the further problem that risks that are not quantified are endangered to be ignored [114:57]. Fairness metrics, for example, may shift the focus of risk assessments and risk management on the composition of data sets and the selection and training of algorithms, possibly neglecting that biases may also emerge during the use of the systems in the interaction between systems and humans.

Eighth, building quantitative measures need consolidated knowledge and consensus in scientific and public discourses. The relevant knowledge base for the risks of applications of AI and ADM is asymmetrically distributed in society, to a large extend in the hands of private providers. This is particularly problematic, because this hinders to replicate, contest, verify or falsify methods, procedures and outcomes of AI and ADM applications. The knowledge is furthermore characterised by a lack of clarity and certainty, in particular, in the evaluative criteria for risk assessments, as sketched out with the normative ambiguities above. Overall, the situation is far from having commonly agreed-upon, inter-subjectively established measurements, which have been replicated, contested, proofed and validated in scientific and public discourses.

## 6.4 Value conflicts

In general, the protection of one fundamental right or societal value may conflict with other fundamental rights or societal values. With regard to AI and ADM, value conflicts seem not to be the exception, but rather ubiquitous. One of the most relevant conflicts is the conflict between the rights to dignity, free development of personality, informational self-determination, privacy and individual justice, on the one hand. On the other hand, there are the rights to freedom to conduct a business and freedom of contract (or contractual autonomy) as well as the societal value of pursuit of efficiency through automation and differentiation of persons.[22] Allowing interference with these fundamental rights, e.g., by accepting certain AI and ADM applications with residual risks of errors or inaccuracies in favour of efficiency arguments, is a normative decision about societal value conflicts. Other intensively discussed and relevant value conflicts are the right to national or public security, which is allegedly improved by extensive use of AI-based surveillance technologies, versus the rights to informational self-determination, data protection and privacy as well as the conflict between transparency and the interest in knowing the decision criteria versus the protection of business and trade secrets.

Many fundamental rights can be balanced against each other, but only under very restrictive conditions. In jurisprudence on fundamental rights, this is usually done, if at all, according to the principles of legality, necessity and proportionality (in particular according to Article 52 Charter). The principles require a legal basis for restrictions on fundamental rights as well as judgments about the importance, advantageousness and effectiveness of an application or measure in achieving a legitimate objective. The judgments about necessity and proportionality also require an examination of whether the application or measure is necessary for the objective, whether the objective is in the public interest, whether no less-intrusive alternatives are available, whether the essence of the impaired right is still respected and whether the restriction of fundamental rights is not disproportionate to the objectives pursued [e.g., 25:52f.; 37:9-12].

---

[22] This trade-off is that of the phenomenon of statistical or rational discrimination with protected attributes or proxies used to evaluate individuals for reasons of efficiency [115; 79].

For risk regulation, such provisions for balancing exercises of conflicting values can provide indications for distinguishing legitimate from illegitimate risk impositions or acceptable from inacceptable risks. This also demonstrates the importance of these decisions in the sense that such comparisons, weightings and derogations of fundamental rights are normally made by highest courts or legislation (see Section 7 how this changes with the proposed AIA).[23] In this context, it should also be noted that the knowledge base for assessing necessity and proportionality is still weak and constantly developing [similar 116]. The benefits of automated decisions using AI methods are not yet fully scientifically proven, such as actual improvements in efficiency, accuracy and 'objectivity' compared to human decisions or compared across different systems, methods and application types.

Furthermore, recent AI and ADM developments often prolong and intensify the impacts of information and communication technology used for decades (e.g., profiling, scoring, data mining or analytics). However, developments in AI and ADM can change tipping points where existing societal agreements on balancing value conflicts break down because the proportionality of encroaching on one right at the expense of another no longer holds. For risk regulation, this means that the conditions for determining a risk as an illegitimate encroachment on a fundamental right are changing. This requires further political procedures to regularly determine the boundaries between legitimate and illegitimate practices based on ever-evolving AI and ADM technologies, i.e. to update the balancing decisions [similar 56:70].

For example, compared to previous data analysis systems, AI systems can infer personality traits, such as character, emotional, psychological or social states or conditions, from seemingly 'harmless' data such as communications on online social networks or from photos, voice or video recordings [e.g., 117; 41:15-16]. This has considerable implications for the balancing of conflicting values. For instance, AI-based biometric recognition technology for public or private applications (e.g., surveillance) can process and analyse extensive personal profiles, can be discriminatory due to biased training data and higher error rates for certain groups of persons (e.g., marginalised groups), be misused to illegitimately infer personality traits that are used for subverting personal decision-making and manipulations or for other illegitimate purposes, process data of a disproportionate high number of persons to identify only a few, and have significant chilling effects on the use of places, events, services or products due to its operation and data use practices that are hardly recognisable for those affected [37; 68:12; 118]. As another example, such AI systems used in marketing, financial or employment contexts can potentially infer personal dependency on products, services or positions such as jobs, credits, or housing, thereby reinforcing the structural dominance of the provider over the persons on the demand side and thus shifting the conditions of legality originally established by regulatory frameworks [41:64f.]. These imbalances can be further accelerated by losses of consumer and citizen choices due to reductions in analogous alternatives or changes in digital markets with larger information asymmetries, deficits of the informed consent approach, market concentration through network effects or larger power asymmetries.

---

[23] For the determination of discrimination, for example, decisions about the acceptable level of bias or thresholds for illegal disparity are traditionally made by national courts or the Court of Justice of the European Union [48:26].

## 6.5   Acceptable risk level

In particular, the determination of acceptable risks levels or the acceptable extent of risks that decision-makers impose on affected individuals in the form of threshold values, criteria or principles, includes normative and political decisions [in general 96; 9:65, 149-156; 12; 119:28-30]. For determining an acceptable risk level in risk regulation of AI and ADM, the basic decisions are whether fundamental rights are to be encroached at all, for which purposes and interests and under which conditions. The proposed AIA has among its objectives the protection of fundamental rights (e.g., Recital 1, 4, 5, or 13 proposed AIA). On the one hand, the AIA proposal aims to achieve their protection by prohibiting certain AI applications. On the other hand, the regulatory requirements for 'high-risk' applications mean that certain degrees of potential restrictions on fundamental rights are hold as acceptable. From this follows a range of normative decisions for the practicalities of risk regulation.

First, the setting of acceptable risk levels in other areas of risk regulation, such as environmental protection or occupational safety and health, can be supported by empirical knowledge about cause-effect relationships and threshold levels of inputs to the environment or exposures in the workplace at which certain forms of harm occur. For the protection of fundamental rights, the definition of harms and threshold values about acceptable encroachments of fundamental rights are mere normative and political decisions. Scientific knowledge, like that about causes of AI-based discriminations, can inform about cause-effect relations but does not inform the definition of threshold values of acceptable risks.

Second, risk assessments and risk management are never perfect and a zero-risk situation is mostly not achievable,[24] neither will they be under the AIA regime. A central normative decision about the acceptable risk level is the definition of the scope of high-risk applications, and thus, the determination of all non-high-risks applications as acceptable. The underlying decision is that between Type I or Type II errors. Here, the question is: "should the regulator lean on the side of individuals with the risk of including non-high risk AI systems in the high risk category (cf., Type

---

[24]   The risk-based regulation requires that regulators takes risks [12:193].

I error), or should they lean on the side of AI system providers and fail to consider an AI system as being high risk when that should have been the case (cf., Type II error)" ([120:28] with reference to [12:188]).[25]

Third, another particularity of the AIA proposal is that the level of risks will actually strongly dependent on the level of resources, time and efforts devoted to testing, identifying, analysing, assessing and managing the risks of AI and ADM applications. Either these decisions will be regulated through common specifications or standardisation (see below), or these decisions fall within the scope of developers or deployers. The decisions include, in particular, the level of accuracy required for risk assessments and how much costs for reaching the level is required to be taken by the regulatees, such as decisions about sample size, the intensity of testing and simulation, or required types of evidences. Further normative decisions include which types of errors (false positives or false negatives) have to be avoided, the appropriate properties of used data sets [37:33f.], the level of risks of adversarial attacks on AI systems or of the risks of misuse and misinterpretation by human operators. Even when society would agree on certain metrics, the setting of thresholds, baselines or target values of acceptable risk levels 'within' these metrics are normative decisions.

Fourth, some human and fundamental rights and fundamental values are founded on human dignity, such as the rights to free development of personality, non-discrimination, informational self-determination, data protection or privacy, freedom of expression and political participation and concepts of justice that emphasise capabilities for a decent life and respect for and recognition of the least advantaged. Here, the meaning of human and fundamental rights derives from the protection of human dignity and its incommensurability and inviolability [23:10; 121; 4:para. 16]. Therefore, risk regulation should involve determining applications and contexts in which fundamental rights must not be violated, trade-offs must be avoided or residual risks are not tolerable. This includes that fundamental rights provide a minimum level of treatment to live a life with dignity and that tolerating residual impacts can mean falling below a minimum level of human dignity. As consequence for risk regulation, the human and fundamental rights approach demands that those residual impacts on human rights are not tolerable and that further remedial measures be taken [122:524]. This is particularly relevant for the 'high-risk' AI application in the AIA proposal. For instance, the European Commission has sorted AI applications for law enforcement into the 'high-risks' category (Annex III (6) proposed AIA). By this, the *de facto* tolerating of error rates in calculating risk scores of offending or reoffending can thus mean to tolerate infringements of the right to human dignity through false incarceration.

---

[25] Gellert [120:29f.] argues that, according to the precautionary principle underlying other forms of European risk regulation, the regulatory scope should also include non-high risks applications.

## 7    Problematic dispersion of normative decisions in the AIA proposal

According to the approach of hybrid governance chosen for the AIA, normative decisions are spread over a number of actors. As main decisions about acceptable risks, the European Commission provides the definition of artificial intelligence and, by this, the scope of the regulation implicitly determining which systems and applications are considered risky to fundamental rights. The Commission establishes and updates the risk categories and the sorting of AI application types to them, and thereby defines systems with unacceptable or acceptable risks. The proposed Act outlines some criteria for the European Commission to be considered when AI systems are assessed and evaluated as high-risk (Article 7 (2) proposed AIA). However, no further guidelines are provided on the interpretation and operationalisation of the criteria, leaving it open how to operationalise "adverse impact on fundamental rights" or "significant concerns" (Article 7 (2) c proposed AIA). Additionally, the national supervisory agencies or market surveillance authorities (MSAs) responsible for the *ex post* market surveillance have to make several normative decisions concerning the criteria to determine when investigative and regulatory actions should be triggered or when the risks of AI and ADM applications are so severe that systems should be corrected, prohibited or withdrawn from the market.

The AIA proposal gives a crucial role of normative decision-making to standardisation bodies and/or developers and providers. For providers of high-risks systems, conformity to the AIA requirements can be assumed when they follow harmonised standards provided by standardisation organisations (Article 40 proposed AIA).[26] The European Commission can mandate the development of harmonised standards to European Standardisation Organisations[27] [124:104f.]. If harmonised standards do not exist or the European Commission find the harmonised standards insufficient, the European Commission can adopt common specifications (Article 41 proposed AIA). In the Impact Assessment accompanying the AIA proposal, the Commission noted that "technical standards may not be fully compliant with existing EU legislation (e.g., on data protection or non-discrimination)" [104:27; with reference to 125].[28]

---

[26]   The risk management measure should also consider the harmonised standards (Article 9 (3) proposed AIA). For initiatives by standardisation organisations (e.g., International Organization for Standardization (ISO), European Committee for Standardization (CEN), European Committee for Electrotechnical Standardization (CENELEC), Deutsches Institut für Normung (DIN)) related to AI see, for instance, [123].

[27]   They are the CEN, CENELEC, and the European Telecommunications Standards Institute (ETSI).

[28]   Christofi and co-authors [125] show that the standardisation of the Privacy Impact Assessment by ISO (ISO/IEC 29134) deviates from the provisions of the GDPR on the Data Protection Impact Assessment.

With respect to fundamental rights, social rule-setting by standardisation bodies is problematic. First, under EU law, the delegation of rule-setting powers to private standardisation bodies raises issues of constitutionality mainly due to the lack of judicial oversight and judicial scrutiny [126:213-214; 124:104-106]. Second, standardisation bodies are no legitimate representatives of society. Their procedures are often less transparent, suffer from asymmetries for different stakeholder groups in resources necessary to participate, and lack a systematic inclusion of stakeholders or persons affected [127; 128; 129]. Furthermore, they have no sufficient democratic control, nor the same procedural safeguards and options of public scrutiny and debate as for legislation. Third, according to Veale and Zuiderveen Borgesius they lack experiences of interpretation and operationalisation of fundamental rights [124:115].

In case the expected standardisation does not provide concrete, substantive specifications of fundamental rights or specifications through balancing between fundamental rights, or deliberately leave discretion to the standards' addressees, those normative decisions fall to the developers or providers of AI systems. The latter are mostly private companies. Therefore, either private standardisation bodies or private providers make profound normative decisions about the interpretation of fundamental rights, about the actual quality and scope of the quality management system and the risk management system[29], about 'appropriate' risks management measures and acceptable levels of risks and residual risks[30], 'appropriate' and 'suitable' levels of and resources for testing[31], or about 'appropriate' levels of accuracy, robustness and cybersecurity required for the design and development of AI systems[32]. This includes also normative decisions not only about which of the metrics are applied, but also about baselines or thresholds 'within' the metrics or measurements, such as error rates imposed on persons affected.

In case the standardisation initiatives focus mainly on process, management or procedural standards (e.g., standards for procedures of risk management), and not on substantive standards about acceptable risks levels etc., standard addressees may tend to focus mainly on mere compliance with standards.[33] Then, it is questionable if providers identify, investigate and assess the potential risks to fundamental rights from the perspective of the persons affected. For instance, it is less likely that providers consider risks to the common good, systemic risks like potential chilling effects, and the distribution of risks among groups of populations. This includes questions of the composition of affected populations like potential impacts on already disadvantaged groups or minorities, whether they should be particularly protected or whether and how social inequalities, possibly perpetuated or aggravated by AI and ADM applications, should be corrected.

---

[29] Article 17 and 9 proposed AIA.
[30] Article 9 (4) proposed AIA.
[31] Article 9 (5) and (6) proposed AIA.
[32] Article 15 (1) proposed AIA.
[33] See for the critique of similar provisions in the GDPR and the problem of 'box-ticking' exercises [125:144, 153].

With the AIA proposal, not courts or legislation, but standardisation bodies, developers, providers and/or (to a smaller extent also) deployers make far-reaching normative decisions about realisations or infringements of fundamental rights through the standardising, design, development, deployment, making the settings on systems, or in framing the 'learning' and adjusting of systems. This situation is inappropriate, because such decisions sensitive to fundamental rights are normally to be made by courts or legislation according to principles of legality, necessity and proportionality [similar 37:29-31, 38]. Under these principles (see above), the proportionality is mostly evaluated whether the encroachment of a fundamental right is justified on the ground that a conflicting objective is given by a pressing social need[34] or it is in the public interest and, thus, benefits the entire society (e.g., national security). In contrast, when such normative balancing decisions are made by developers or providers they are made by balancing violations of fundamental rights with economic interests in terms of profits, benefiting only their own interest. This is a huge difference in the application of the proportionality test, that is particularly problematic under the circumstance that the essence of fundamental rights is not clear or is in need to be adapted to new socio-technological developments. Furthermore, balancing decisions by developers, providers or deployers lack legal certainty either because they detect by themselves that the aims can be achieved by less adverse alternatives and have to refrain from development and deployments (as normally required by the principles) or the application's use may be stopped through later revisions by supervisory authorities or court decisions.

The inappropriateness is all the more relevant in the light of the severity of potential violations of fundamental rights and the above-mentioned 'mass phenomenon' of ADM applications. Such normative choices affect broad segments of the population. For instance, systems with residual error rates can easily discriminate large segments of the population. This scale effect is especially critical for applications where affected persons have no alternatives for evasion such as in public services, applications by state authorities or on concentrated markets. In view of this scale effect, it would be more appropriate when major normative decisions of risk regulation take place in inclusive democratic processes involving the people actually or potentially affected and their elected policy-makers [see also 61].

---

[34] See for the difference of the proportionality tests in the rights-based and risk-based approaches in data protection law [17:9-24].

The situation is particularly inappropriate regarding the normative ambiguities in evaluative criteria that make the discretion in risk assessment and risk management even greater and may promote arbitrariness. In general, risk governance when designed as meta-regulation includes the delegation of risk-regulatory task to diverse governance actors including private actors or regulatees respectively. The appropriateness of this governance options depends, among other things, on whether objective or understandable knowledge is generated by providers that can be verified by outsiders such as the regulator [similar 111:420-422]. However, as long as normative ambiguities prevail so that risk assessments are not objective and are not based on scientific and public consensus, such delegations hardly seem suitable. Risk assessment by the providers cannot be considered as objective scientific endeavour, but as context-dependent subjective balancing exercising of trade-offs.

Furthermore, science and technology studies on risk perception have shown that the interpretation of risks facts is influenced by contextual information [e.g., 97:127f.]. Thus, the outcomes of risk assessment and risk management dependent on factors specific to the different providers of AI systems, or risk assessors respectively, such as the individual risk appetite, experiences, knowledge about fundamental rights or insights about or attitudes towards the potentially affected population. This leads to incongruent or fragmented levels of protection of fundamental rights. Fundamental rights are, however, universal in the sense that all people equally have fundamental rights.

## 8    Conclusions

We have shown that several normative ambiguities in risk regulation of AI and ADM applications exist. They necessitate normative decisions on establishing and maintaining a risk regulation scheme. These choices include decisions about the specific form of the risk-based approach itself, the selection and prioritisation of risks, the choice of conceptions underlying interpretations and operationalisations of fundamental rights and values, choices of approaches to aggregate or quantify risks, the balancing of value conflicts, the updating of balancing decisions, or the determination of acceptable risk levels. Since normative choices normally belong to devising and running a risk regulation scheme, it is a question of who makes those decisions and in whose interest.

Decisions about trade-offs of fundamental rights among each other or against other fundamental values play a central role to determine acceptable risk types and levels, and they are one of the sources of normative ambiguities, which are usually prone to conflicts. The challenges of value conflicts for risk regulation include that value conflicts are not recognised as such and as a normative matter of risk regulation. Trade-offs of risks can lead to residual risks that can be actually unacceptable with regard to human and fundamental rights protection. Furthermore, trade-offs might be decided in view of an outdated and inadequate balancing institutionalised in secondary legislation. Further problems are that trade-offs are made by actors who lack democratic legitimacy. They can also be based on an inadequate formal foundation (e.g., without a legal basis) or are done concealed although the procedures should actually take place in public political processes. Under these conditions, a further problem might be a distortion of balancing in favour of particular private interests, disregard public interests such as the overall distribution of risks and benefits.

The hybrid governance approach of the proposed AIA results in a problematic diffusion of normative decision-making across several actors. Besides the European Commission, national supervisory authorities and market surveillance authorities, crucial normative decisions will be made by standardisation organisations and/or the developers and providers of AI and ADM systems. Standardisation organisations and providers lack the legitimacy, options of democratic control, and competence to decide on encroachments of fundamental rights. This is also inappropriate with respect to the society-wide scope of their impacts on fundamental rights that AI and ADM applications potentially can have. AI-based ADM systems can easily affect large portions of a population. The risk-based approach in the form chosen for the AIA can endanger that fundamental rights are guaranteed for everyone.

Decisions about balancing between fundamental rights and their limitations are normally made by highest courts or legislation. Court decisions also have their shortcomings as source of specification of fundamental rights and balancing of trade-offs as they relate to the specific situations of the case and transfers of meaning to other contexts are not always possible. Furthermore, in order to bring about court decisions, there must be a plaintiff that files a suit. Thus, legal clarification and operationalisation of fundamental rights with court decisions could be fragmented and last for years. Instead, substantial regulatory standards as result of legislation have the advantages to give clear guidance for developers and providers on legitimate development paths and enables legal security of investments.

Normative decisions inherent in risk regulation should be identified as such and delegated to legitimate, democratic political processes[35] supported by knowledge from systematic research and publicly discussed evidence. Because risks to fundamental rights are imposed on those affected by AI and ADM applications, these normative choices should be negotiated in public discourses [34:103f.; 96:136f.]. This should be done before risk regulation schemes are established and regulatory tasks are distributed among governance actors in order to prevent ambiguities, and failures in their conduct, to provide sufficient guidance to regulators and regulatees and, ultimately, to ensure legitimacy of the regulated AI and ADM applications and their impacts on fundamental rights and societal values. A 'hidden privatisation' of decisions about public values [19:34f.] and value conflicts must be avoided.

---

[35] However, the proposed AIA does not establish rights of consultation and public participation for stakeholders [37:48-54]. Furthermore, in this respect, it remains to be seen whether the AIA provisions on common specifications (Article 3 (28) and 41 proposed AIA) are used and how democratically the related political processes are shaped.

# 9 References

1. European Parliament (2020) Resolution of 20 October 2020 with recommendations to the Commission on a framework of ethical aspects of artificial intelligence, robotics and related technologies (2020/2012(INL)). European Parliament, Strasbourg
2. European Commission (2020) White Paper On Artificial Intelligence - A European approach to excellence and trust. COM(2020) 65 final. European Commission, Brussels
3. Council of Europe - CAHAI (2020) Feasibility Study, CAHAI (2020)23 Ad hoc Committee on Artificial Intelligence. Council of Europe, Ad Hoc Committee on Artificial Intelligence (CAHAI), Strasbourg
4. Council of Europe - CAHAI (2021) Possible elements of a legal framework on artificial intelligence, based on the Council of Europe's standards on human rights, democracy and the rule of law. CM(2021)173-add, 17 Dec 2021. Council of Europe - Ad hoc Committee on Artificial Intelligence (CAHAI), Strasbourg
5. Council of Europe - European Committee of Ministers (2020) Recommendation CM/Rec(2020)1 of the Committee of Ministers to Member States on the human rights impacts of algorithmic systems. Council of Europe, Strasbourg
6. German Data Ethics Commission (2019) Opinion of the Data Ethics Commission. Data Ethics Commission of the Federal Government; Federal Ministry of the Interior, Building and Community; Federal Ministry of Justice and Consumer Protection, Berlin
7. Enquete-Kommission Künstliche Intelligenz (2020) Bericht der Enquete-Kommission Künstliche Intelligenz - Gesellschaftliche Verantwortung und wirtschaftliche, soziale und ökologische Potenziale (Drucksache 19/23700, 28.10.2020). Deutscher Bundestag, Berlin
8. Hood C, Rothstein H, Baldwin R (2001) The Government of Risk: Understanding Risk Regulation Regimes. Oxford University Press, Oxford
9. Renn O (2008) Risk Governance. Coping with Uncertainty in a Complex World. Earthscan, London
10. Black J (2010) The Role of Risk in Regulatory Processes. In: Baldwin, R, Cave, M, Lodge, M (eds.): The Oxford Handbook of Regulation. Oxford University Press, Oxford, pp. 302-348
11. van der Heijden J (2019) Risk governance and risk-based regulation: A review of the international academic literature, State of the Art in Regulatory Governance Research Paper Series. Victoria University of Wellington, Wellington
12. Black J (2010) Risk-based Regulation: Choices, Practices and Lessons Being Learnt. In: OECD (ed.): Risk and Regulatory Policy: Improving the Governance of Risk, OECD Reviews of Regulatory Reform. Organisation for Economic Co-Operation and Development (OECD), Paris, pp. 185-236

13. Macenaite M (2017) The "Riskification" of European Data Protection Law through a two-fold Shift. Eur J Risk Regul (8)3: 506-540, https://doi.org/10.1017/err.2017.40
14. Hutter BM (2005) The Attractions of Risk-based Regulation: accounting for the emergence of risk ideas in regulation. ESRC Centre for Analysis of Risk and Regulation, London
15. Rothstein H, Irving P, Walden T, Yearsley R (2006) The risks of risk-based regulation: Insights from the environmental policy domain. Environ Int (32)8: 1056-1065, https://doi.org/10.1016/j.envint.2006.06.008
16. Black J, Baldwin R (2010) Really responsive risk-based regulation. Law Pol (32)2: 181-213, https://doi.org/10.1111/j.1467-9930.2010.00318.x
17. Gellert R (2020) The Risk-Based Approach to Data Protection. Oxford University Press, Oxford
18. Zuiderveen Borgesius F (2018) Discrimination, Artificial Intelligence, and Algorithmic Decision-Making. Council of Europe, Directorate General of Democracy, Strasbourg
19. Yeung K (2019) Responsibility and AI. A study of the implications of advanced digital technologies (including AI systems) for the concept of responsibility within a human rights framework. Council of Europe study DGI(2019)05. Council of Europe, Expert Committee on human rights dimensions of automated data processing and different forms of artificial intelligence (MSI-AUT), Strasbourg
20. Helberger N, Eskens S, van Drunen M, Bastian M, Moeller J (2020) Implications of AI-Driven Tools in the Media for Freedom of Expression. Council of Europe, Strasbourg
21. Wagner B (2018) Algorithms and Human Rights. Study on the human rights dimensions of automated data processing techniques and possible regulatory implications. Council of Europe study DGI(2017)12. Council of Europe, Committee of experts on internet intermediaries (MSI-NET), Strasbourg
22. Mantelero A (2019) Artificial Intelligence and Data Protection: Challenges and Possible Remedies. Consultative Committee of the Convention for the Protection of Individuals with Regard to Automatic Processing of Personal Data (Convention 108), Report on Artificial Intelligence. Council of Europe, Directorate General of Human Rights and Rule of Law, Strasbourg
23. AI HLEG (2019) Ethics Guidelines for Trustworthy AI. Independent High-Level Expert Group on Artificial Intelligence (AI HLEG), report published by the European Commission, Brussels
24. AI HLEG (2019) Policy and Investment Recommendations for Trustworthy AI. Independent High-Level Expert Group on Artificial Intelligence (AI HLEG), report published by the European Commission, Brussels
25. FRA (2020) Getting the Future Right - Artificial Intelligence and Fundamental Rights. European Union Agency for Fundamental Rights (FRA), Publications Office of the European Union, Luxembourg
26. Access Now (2018) Human Rights in the Age of Artificial Intelligence. Access Now, Brooklyn
27. Latonero M (2018) Governing Artificial Intelligence: Upholding Human Rights and Dignity. Data & Society Research Institute, New York
28. Raso FA, Hilligoss H, Krishnamurthy V, Bavitz C, Kim L (2018) Artificial Intelligence & Human Rights: Opportunities & Risks. Berkman Klein Center for Internet & Society, Cambridge, MA
29. Donahoe E, MacDuffee Metzger M (2019) Artificial Intelligence and Human Rights. J Democr (30)2: 115-126, https://doi.org/10.1353/jod.2019.0029
30. Mantelero A, Esposito MS (2021) An evidence-based methodology for human rights impact assessment (HRIA) in the development of AI data-intensive systems. Comput Law Secur Rev (41): 105561, https://doi.org/10.1016/j.clsr.2021.105561

31. UN (2018) Report on Artificial Intelligence technologies and implications for freedom of expression and the information environment. Report of the Special Rapporteur on the promotion and protection of the right to freedom of opinion and expression, David Kaye. Report A/73/348. United Nations, OHCHR, Geneva
32. Bundesregierung (2018) Strategie Künstliche Intelligenz der Bundesregierung. Bundesregierung, Berlin
33. Nickel JW (2007) Making Sense of Human Rights. Wiley-Blackwell, Malden, Oxford
34. Shrader-Frechette K (2005) Flawed attacks on contemporary human rights: Laudan, Sunstein, and the cost-benefit state. Hum Rights Rev (7)1: 92-110, https://doi.org/10.1007/s12142-005-1004-1
35. Nemitz P (2018) Constitutional democracy and technology in the age of artificial intelligence. Philos Trans A Math Phys Eng Sci (376)2133: 1-14, https://doi.org/10.1098/rsta.2018.0089
36. Smuha NA (2020) Beyond a Human Rights-Based Approach to AI Governance: Promise, Pitfalls, Plea. Philos Technol (34)Suppl. iss. 1: 91-104, https://doi.org/10.1007/s13347-020-00403-w
37. Smuha NA, Ahmed-Rengers E, Harkens A, Li W, MacLaren J, Piselli R, Yeung K (2021) How the EU Can Achieve Legally Trustworthy AI: A Response to the European Commission's Proposal for an Artificial Intelligence Act. LEADS Lab, University of Birmingham, Birmingham
38. European Commission (2018) Communication (2018) 237 final on Artificial Intelligence in Europe. European Commission, Brussels
39. AI HLEG (2018) A Definition of AI: Main Capabilities and Disciplines. Independent High-Level Expert Group on Artificial Intelligence (AI HLEG), report published by the European Commission, Brussels
40. Yeung K (2019) Why Worry about Decision-Making by Machine? In: Yeung, K, Lodge, M (eds.): Algorithmic Regulation. Oxford University Press, Oxford, pp. 21-48
41. Orwat C (2020) Risks of Discrimination through the Use of Algorithms. A study compiled with a grant from the Federal Anti-Discrimination Agency. Federal Anti-Discrimination Agency, Berlin
42. European Commission (2020) Report on the safety and liability implications of Artificial Intelligence, the Internet of Things and robotics. European Commission, Brussels
43. Kleinberg J, Ludwig J, Mullainathan S, Sunstein CR (2018) Discrimination in the Age of Algorithms. J Leg Anal (10): 113-174, https://doi.org/10.1093/jla/laz001
44. Burrell J (2016) How the machine 'thinks': Understanding opacity in machine learning algorithms. Big Data Soc (3)1: 1-12, https://doi.org/10.1177%2F2053951715622512
45. Brkan M, Bonnet G (2020) Legal and Technical Feasibility of the GDPR's Quest for Explanation of Algorithmic Decisions: of Black Boxes, White Boxes and Fata Morganas. Eur J Risk Regul (11)1: 18-50, https://doi.org/10.1017/err.2020.10
46. Kroll JA (2018) The fallacy of inscrutability. Philos Trans A Math Phys Eng Sci (376)2133: 1-14, https://doi.org/10.1098/rsta.2018.0084
47. Hacker P (2018) Teaching Fairness to Artificial Intelligence: Existing and Novel Strategies Against Algorithmic Discrimination Under EU Law. Common Mark Law Rev (55)4: 1143-1185
48. Wachter S, Mittelstadt B, Russell C (2021) Why fairness cannot be automated: Bridging the gap between EU non-discrimination law and AI. Comput Law Secur Rev (41): 105567, https://doi.org/10.1016/j.clsr.2021.105567
49. Jobin A, Ienca M, Vayena E (2019) The global landscape of AI ethics guidelines. Nat Mach Intell (1)9: 389-399, https://doi.org/10.1038/s42256-019-0088-2

50. Fjeld J, Achten N, Hilligoss H, Nagy A, Srikumar M (2020) Principled Artificial Intelligence: Mapping Consensus in Ethical and Rights-Based Approaches to Principles for AI. Berkman Klein Center Research Publication No. 2020-1. Berkman Klein Center for Internet & Society at Harvard University, Cambridge, MA
51. Rudschies C, Schneider I, Simon J (2021) Value Pluralism in the AI Ethics Debate – Different Actors, Different Priorities. Int J o Inf Ethics (29)3: 1-15, https://doi.org/10.29173/irie419.
52. Wagner B (2018) Ethics as an escape from regulation. From "ethics-washing" to ethics-shopping? In: Bayamlioğlu, E, et al. (eds.): Being Profiled: Cogitas Ergo Sum. 10 Years of 'Profiling the European Citizen'. Amsterdam University Press, Amsterdam, pp. 84-88
53. Deutscher Ethikrat (2018) Big Data und Gesundheit – Datensouveränität als informationelle Freiheitsgestaltung. Stellungnahme. Deutscher Ethikrat, Berlin
54. Hildebrandt M (2016) Law as Information in the Era of Data-Driven Agency. Mod Law Rev (79)1: 1-30, https://doi.org/10.1111/1468-2230.12165
55. Ruggiu D (2018) Human Rights and Emerging Technologies. Analysis and Perspectives in Europe. Pan Stanford, New York
56. Ruggiu D (2019) Models of Anticipation Within the Responsible Research and Innovation Framework: the Two RRI Approaches and the Challenge of Human Rights. NanoEthics (13)1: 53-78, https://doi.org/10.1007/s11569-019-00337-4
57. Yeung K, Howes A, Pogrebna G (2020) AI Governance by Human Rights-Centred Design, Deliberation and Oversight: An End to Ethics Washing. In: Dubber, M, Pasquale, F, Das, S (eds.): The Oxford Handbook of AI Ethics. Oxford University Press, New York, pp. 77-106
58. Götzmann N, Vanclay F, Seier F (2016) Social and human rights impact assessments: what can they learn from each other? Impact Assess Proj Apprais (34)1: 14-23, https://doi.org/10.1080/14615517.2015.1096036
59. Johansen IL, Rausand M (2015) Ambiguity in risk assessment. Saf Sci (80): 243-251, https://doi.org/10.1016/j.ssci.2015.07.028
60. Stirling A (2008) Science, Precaution, and the Politics of Technological Risk. Ann NY Acad Sci (1128)1: 95-110, https://doi.org/10.1196/annals.1399.011
61. EGE (2018) Statement on Artificial Intelligence, Robotics and 'Autonomous' Systems, European Group on Ethics in Science and New Technologies (EGE), European Commission Brussels
62. Jones ML (2017) The right to a human in the loop: Political constructions of computer automation and personhood. Soc Stud Sci (47)2: 216-239, https://doi.org/10.1177/0306312717699716
63. McCrudden C (2008) Human dignity and judicial interpretation of human rights. Eur J Int Law (19)4: 655-724, https://doi.org/10.1093/ejil/chn059
64. Becchi P, Mathis K (2019) Handbook of Human Dignity in Europe. Springer International Publishing, Cham
65. Schaber P (2013) Instrumentalisierung und Menschenwürde. Mentis, Münster
66. Düwell M (2017) Human Dignity and the Ethics and Regulation of Technology. In: Brownsword, R, Scotford, E, Yeung, K (eds.): The Oxford Handbook of Law, Regulation and Technology. Oxford University Press, Oxford, pp. 177-196
67. Schaber P (2012) Menschenwürde. Reclam, Stuttgart

68. EDPB/EDPS (2021) Joint Opinion 5/2021 on the proposal for a Regulation of the European Parliament and of the Council laying down harmonised rules on artificial intelligence (Artificial Intelligence Act). European Data Protection Board (EDPB), European Data Protection Supervisor (EDPS), Brussels

69. Scholz P (2019) DSGVO Art. 22 Automatisierte Entscheidungen im Einzelfall einschließlich Profiling. In: Simitis, S, Hornung, G, Spiecker genannt Döhmann, I (eds.): Datenschutzrecht. DSGVO und BDSG. Nomos, Baden-Baden

70. de Terwangne C (2022) Privacy and data protection in Europe: Council of Europe's Convention 108+ and the European Union's GDPR. In: Fuster, GG, Van Brakel, R, Hert, Pd (eds.): Research Handbook on Privacy and Data Protection Law. Values, Norms and Global Politics, Research Handbooks in Information Law series. Edward Elgar, Cheltenham, Northampton, pp. 10-35

71. Brkan M (2019) The Essence of the Fundamental Rights to Privacy and Data Protection: Finding the Way Through the Maze of the CJEU's Constitutional Reasoning. Ger Law J (20)6: 864-883, https://doi.org/10.1017/glj.2019.66

72. Fuster GG, Gutwirth S (2013) Opening up personal data protection: A conceptual controversy. Comput Law Secur Rev (29)5: 531-539, https://doi.org/10.1016/j.clsr.2013.07.008

73. Britz G (2010) Informationelle Selbstbestimmung zwischen rechtswissenschaftlicher Grundsatzkritik und Beharren des Bundesverfassungsgerichts. In: Hoffmann-Riem, W (ed.): Offene Rechtswissenschaft. Mohr Siebeck, Tübingen, pp. 561-596

74. Rouvroy A, Poullet Y (2009) The Right to Informational Self-Determination and the Value of Self-Development: Reassessing the Importance of Privacy for Democracy. In: Gutwirth, S, et al. (eds.): Reinventing Data Protection? Springer, Amsterdam, pp. 45-76

75. Solove DJ (2006) A Taxonomy of Privacy. U Pa Law Rev (154)3: 477-560, https://doi.org/10.2307/40041279

76. Tavani HT (2008) Informational privacy: Concepts, theories, and controversies. In: Himma, KE, Tavani, HT (eds.): The Handbook of Information and Computer Ethics. John Wiley and Sons, Hoboken, NJ, pp. 131-164

77. Koops B-J, Newell BC, Timan T, Skorvanek I, Chokrevski T, Galic M (2016) A typology of privacy. U Pa J Int Law (38)2: 483-575

78. Drackert S (2014) Die Risiken der Verarbeitung personenbezogener Daten. Eine Untersuchung zu den Grundlagen des Datenschutzrechts. Dunker & Humblot, Berlin

79. Britz G (2008) Einzelfallgerechtigkeit versus Generalisierung. Verfassungsrechtliche Grenzen statistischer Diskriminierung. Mohr Siebeck, Tübingen

80. Wagner I, Eckhoff D (2018) Technical privacy metrics: a systematic survey. ACM Comput Surv (51)3: 1-38, https://doi.org/10.1145/3168389

81. Pohle J (2020) On Measuring Fundamental Rights Protection: Can and Should Data Protection Law Learn From Environmental Law? In: The Global Constitutionalism and the Internet Working Group (ed.): Don't Give Up, Stay Idealistic and Try To Make the World a Better Place - Liber Amicorum for Ingolf Pernice. HIIG, Berlin, pp. 71-79

82. Romei A, Ruggieri S (2014) A multidisciplinary survey on discrimination analysis. Knowl Eng Rev (29)5: 582-638, https://doi.org/10.1017/s0269888913000039

83. Barocas S, Selbst AD (2016) Big Data's Disparate Impact. Cal L Rev (104)3: 671-732, https://doi.org/10.15779/Z38bg31

84. Mehrabi N, Morstatter F, Saxena N, Lerman K, Galstyan A (2021) A survey on bias and fairness in machine learning. ACM Comput Surv (54)6: 1-35, https://doi.org/10.1145/3457607
85. Verma S, Rubin J (2018) Fairness definitions explained, presented at: 2018 IEEE/ACM Int. Workshop on Software Fairness (FairWare); published by IEEE
86. Binns R (2018) Fairness in Machine Learning: Lessons from Political Philosophy, presented at: Conference on Fairness, Accountability, and Transparency (FAT) 2018
87. Mulligan DK, Kroll JA, Kohli N, Wong RY (2019) This Thing Called Fairness: Disciplinary Confusion Realizing a Value in Technology. Proceedings of the ACM on Human-Computer Interaction (3)Article No. 119: 1-36, https://doi.org/10.1145/3359221
88. Hauer MP, Kevekordes J, Haeri MA (2021) Legal perspective on possible fairness measures - Why AI in HR needs help. Comput Law Secur Rev (42)Sept.: 105583, https://doi.org/10.1016/j.clsr.2021.105583
89. Wachter S, Mittelstadt B, Russell C (2020) Bias preservation in machine learning: the legality of fairness metrics under EU non-discrimination law. W Va L Rev (123)3: 735-790
90. Binns R (2020) On the Apparent Conflict Between Individual and Group Fairness, presented at: FAT* '20, January 27–30, 2020, Barcelona, Spain
91. YVTltk (2018) Assessment of creditworthiness, authority, direct multiple discrimination, gender, language, age, place of residence, financial reasons, conditional fine. Plenary Session (voting), Register number: 216/2017, 21 March 2018. Yhdenvertaisuus- ja tasa-arvolautakunta / National Non-Discrimination and Equality Tribunal of Finland, Finland, Government Publication
92. Snelling J, McMillan J (2017) Equality: Old Debates, New Technologies. In: Brownsword, R, Scotford, E, Yeung, K (eds.): The Oxford Handbook of Law, Regulation and Technology. Oxford University Press, Oxford, pp. 69-89
93. Hussain W (2018) The Common Good. In: Zalta, EN (ed.): Stanford Encyclopedia of Philosophy (Spring 2018 Edition)
94. Jaume-Palasi L (2019) Why We Are Failing to Understand the Societal Impact of Artificial Intelligence. Soc Res (86)2: 477-498, https://doi.org/10.1353/sor.2019.0023
95. Mantelero A (2016) Personal data for decisional purposes in the age of analytics: From an individual to a collective dimension of data protection. Comput Law Secur Rev (32)2: 238-255, https://doi.org/10.1016/j.clsr.2016.01.014
96. Fischhoff B, Watson SR, Hope C (1984) Defining Risk. Pol Sci (17)2: 123-139
97. Jasanoff S (1993) Bridging the two cultures of risk analysis. Risk Anal (13)2: 123-123, https://doi.org/10.1111/j.1539-6924.1993.tb01057.x
98. Horlick-Jones T (1998) Meaning and contextualisation in risk assessment. Reliab Eng Syst Saf (59)1: 79-89, https://doi.org/10.1016/S0951-8320(97)00122-1
99. Jasanoff S (1999) The Songlines of Risk. Environ Values (8)2: 135-152, https://doi.org/10.3197/096327199129341761
100. Felt U, Wynne B, Callon M, Gonçalves ME, Jasanoff S, Jepsen M, Joly P-B, Konopasek Z, May S, Neubauer C, Rip A, Siune K, Stirling A, Tallacchini M (2007) Taking European Knowledge Society Seriously. Report of the Expert Group on Science and Governance to the Science, Economy and Society Directorate, Directorate-General for Research, European Commission. Office for Official Publications of the European Communities, Luxembourg
101. Baldwin R, Black J (2016) Driving Priorities in Risk-based Regulation: What's the Problem? J Law Soc (43)4: 565-595, https://doi.org/10.1111/jols.12003

102. Cranor CF (1997) The Normative Nature of Risk Assessment: Features and Possibilities. Risk Health Saf Environ (8)Spring: 123-136
103. Hansson SO (2010) Risk: Objective or Subjective, Facts or Values. J Risk Res (13)2: 231-238, https://doi.org/10.1080/13669870903126226
104. European Commission (2021) Impact Assessment Accompanying the Proposal for a Regulation of the European Parliament and of the Council Laying Down Harmonised Rules on Artificial Intelligence (Artificial Intelligence Act) and Amending Certain Union Legislative Acts; 24.4.2021, SWD(2021) 84 final. European Commission, Brussels
105. Ansell C, Baur P (2018) Explaining Trends in Risk Governance: How Problem Definitions Underpin Risk Regimes. Risk, Hazards Crisis Public Policy (9)4: 397-430, https://doi.org/10.1002/rhc3.12153
106. Sovrano F, Sapienza S, Palmirani M, Vitali F (2022) Metrics, Explainability and the European AI Act Proposal. J - Multidiscipl Sci J (5)1: 126-138, https://doi.org/10.3390/j5010010
107. Goodfellow I, Bengio Y, Courville A (2016) Deep learning. MIT Press, Cambridge
108. Merry SE (2016) The Seductions of Quantification: Measuring Human Rights, Gender Violence, and Sex Trafficking. University of Chicago Press, Chicago, London
109. Bowker GC, Star SL (2000) Sorting Things Out: Classification and Its Consequences. MIT Press, Cambridge, London
110. Mennicken A, Espeland WN (2019) What's New with Numbers? Sociological Approaches to the Study of Quantification. Annu Rev Sociol (45)1: 223-245, https://doi.org/10.1146/annurev-soc-073117-041343
111. Fisher E (2012) Risk and Governance. In: Levi-Faur, D (ed.): Oxford Handbook of Governance. Oxford University Press, Oxford, pp. 417-428
112. Jasanoff S (2004) The idiom of co-production. In: Jasanoff, S (ed.): States of Knowledge. The co-production of science and political order. Routledge, London, New York, pp. 1-12
113. Yeung K, Bygrave LA (2021) Demystifying the modernized European data protection regime: Cross-disciplinary insights from legal and regulatory governance scholarship. Regul Gov (16)1: 137-155, https://doi.org/10.1111/rego.12401
114. Fisher E (2010) Risk Regulatory Concepts and the Law. In: OECD (ed.): Risk and Regulatory Policy: Improving the Governance of Risk, OECD Reviews of Regulatory Reform. OECD, Paris, pp. 45-92
115. Gandy Jr. OH (2009) Coming to Terms with Chance: Engaging Rational Discrimination and Cumulative Disadvantage. Ashgate, Farnham, Burlington
116. Nordström M (2021) AI under great uncertainty: implications and decision strategies for public policy. AI Soc, https://doi.org/10.1007/s00146-021-01263-4
117. Matz SC, Appel RE, Kosinski M (2019) Privacy in the age of psychological targeting. Curr Opin Psychol (31)Feb.: 116-121, https://doi.org/10.1016/j.copsyc.2019.08.010
118. FRA (2019) Facial recognition technology: fundamental rights considerations in the context of law enforcement. European Union Agency for Fundamental Rights, Vienna
119. Grunwald A (2019) Technology Assessment in Theory and Practice. Routledge, London, New York
120. Gellert R (2021) The role of the risk-based approach in the General data protection Regulation and in the European Commission's proposed Artificial Intelligence Act: Business as usual? J Ethics Leg Technol (3)2: 15-33, http://dx.doi.org/10.14658/pupj-jelt-2021-2-2
121. Habermas J (2010) The Concept of Human Dignity and the Realistic Utopia of Human Rights. Metaphilosophy (41)4: 464-480, https://doi.org/10.1111/j.1467-9973.2010.01648.x

122. Mares R (2019) Securing human rights through risk-management methods: Breakthrough or misalignment? Leiden J Int Law (32)3: 517-535, https://doi.org/10.1017/S0922156519000244
123. Nativi S, De Nigris S (2021) AI Standardisation Landscape: state of play and link to the EC proposal for an AI regulatory framework (Report EUR 30772 EN), European Commission, Joint Research Centre. Publications Office of the European Union, Luxembourg
124. Veale M, Zuiderveen Borgesius F (2021) Demystifying the Draft EU Artificial Intelligence Act - Analysing the good, the bad, and the unclear elements of the proposed approach. Comput Law Rev Int (22)4: 97-112, https://doi.org/10.9785/cri-2021-220402
125. Christofi A, Dewitte P, Ducuing C, Valcke P (2020) Erosion by Standardisation: Is ISO/IEC 29134:2017 on Privacy Impact Assessment Up to (GDPR) Standard? In: Tzanou, M (ed.): Personal Data Protection and Legal Developments in the European Union. IGI Global, Hershey, pp. 140-167
126. Van Cleynenbreugel P (2021) EU By-Design Regulation in the Algorithmic Society. A Promising Way Forward or Constitutional Nightmare in the Making? In: Micklitz, H-W, et al. (eds.): Constitutional Challenges in the Algorithmic Society. Cambridge University Press, Cambridge, pp. 202-218
127. Feng P (2006) Shaping Technical Standards. Where Are the Users? In: Guston, DH, Sarewitz, DR (eds.): Shaping Science and Technology Policy: the Next Generation of Research, Science and Technology in Society. University of Wisconsin Press, Madison, pp. 199-216
128. Werle R, Iversen EJ (2006) Promoting Legitimacy in Technical Standardization. Sci Technol Inno Stud (2)2: 19-39
129. Wickson F, Forsberg E-M (2015) Standardising responsibility? The significance of interstitial spaces. Sci Eng Ethics (21)5: 1159-1180, https://doi.org/10.1007/s11948-014-9602-4